\documentclass[aps,prl,superscriptaddress,showpacs,twocolumn]{revtex4-1}
\usepackage{amsfonts}
\usepackage{amsmath}
\usepackage{amssymb}
\usepackage{graphicx}
\usepackage{xcolor}
\providecommand{\U}[1]{\protect\rule{.1in}{.1in}}

\newcommand{\deltaj}{\delta_j}

\begin{document}
\title{Long distance coherence of Majorana wires}
\date{\today}
\author{Zheng Shi}
\affiliation{Dahlem Center for Complex Quantum Systems and Physics Department, Freie Universit\"{a}t Berlin, Arnimallee 14, 14195 Berlin, Germany}
\author{Piet W. Brouwer}
\affiliation{Dahlem Center for Complex Quantum Systems and Physics Department, Freie Universit\"{a}t Berlin, Arnimallee 14, 14195 Berlin, Germany}
\author{Karsten Flensberg}
\affiliation{Dahlem Center for Complex Quantum Systems and Physics Department, Freie Universit\"{a}t Berlin, Arnimallee 14, 14195 Berlin, Germany}
\affiliation{Center for Quantum Devices, Niels Bohr Institute, University of Copenhagen, DK-2100 Copenhagen, Denmark}
\author{Leonid I. Glazman}
\affiliation{Department of Physics, Yale University, New Haven, CT 06520, USA}
\author{Felix von Oppen}
\affiliation{Dahlem Center for Complex Quantum Systems and Physics Department, Freie Universit\"{a}t Berlin, Arnimallee 14, 14195 Berlin, Germany}

\begin{abstract}
Theoretically, a pair of Majorana bound states in a topological superconductor forms a single fermionic level even at large separations, implying that the parity information is stored nonlocally. The nonlocality leads to a long-distance coherence for electrons tunneling through a Coulomb blockaded Majorana wire [Fu, Phys. Rev. Lett. \textbf{104}, 056402 (2010)], an effect that can be observed, {\em e.g.}, in an interferometer. Here, we examine theoretically the coherent electron transfer, taking into account that tunneling implies the long-distance transfer of charge, which is carried by one-dimensional plasmons. We show that the charge dynamics does not affect the coherence of the electron tunneling process in a topological superconductor consisting of a semiconductor wire proximitized by a single bulk superconductor. The coherence may be strongly suppressed, however, if the topological superconductivity derives from a semiconductor wire proximitized by a granular superconductor.
\end{abstract}
\maketitle

\emph{Introduction.---}
One-dimensional topological superconductors have Majorana
bound states (MBSs) localized at their boundaries. The MBSs at both ends
together form a single, highly nonlocal fermionic level, carrying information
about the total fermion parity of the macroscopic superconductor. This
property is part of the basis for the ideas to use MBSs for topologically
protected quantum
computation \cite{RevModPhys.80.1083,PhysRevB.88.035121,PhysRevX.5.041038,PhysRevB.94.174514,PhysRevB.95.235305,PhysRevB.94.235446,NewJPhys.19.012001,PhysRevX.7.031048,PhysRevB.97.205404}.

The nonlocality leads to a striking long-distance coherence when Coulomb interactions are included. This was pointed out by Fu \cite{PhysRevLett.104.056402}, who argued that, at energies smaller than the superconducting gap and the charging energy, a Coulomb blockaded topological superconductor wire is equivalent to a fermionic level with support at the two ends of the wire. Consequently, single electron coherence should be observable at distances far beyond the coherence length of the superconductor, for instance in an Aharonov-Bohm interferometer \cite{PhysRevLett.104.056402}. This long-distance coherence has been suggested as a way to test whether experimentally observed zero-bias peaks \cite{NatRevMater.3.52} originate from isolated MBSs or localized Andreev bound states \cite{PhysRevB.98.161401,PhysRevLett.122.117001,PhysRevB.97.161401}. The first experiments in this direction have already been done, and seem consistent with the long-distance coherence picture \cite{1902.07085}.

In Fu's original derivation \cite{PhysRevLett.104.056402}, it is assumed that the electron and its charge is instantaneously distributed in the wire. This is usually justified by noting that the charge rearrangement in metals happens on the short timescale of the inverse plasma frequency. However, this need not a priori apply to a wire geometry, where the charge modes are one-dimensional (surface) plasmons with a linear dispersion.

In this paper, we present a theory of the long-distance coherent transport through a topological superconductor that includes charge redistribution effects. Specifically, we consider a system consisting of two Majorana wires in an interference loop setup. The wires are assumed to have a bulk excitation gap and to be much longer than the superconducting coherence length, so that no subgap Andreev states extend from one end to the other. Hence, the only mechanism for (subgap) coherent transfer of single electrons is via the end MBSs. Our theoretical description of this effect takes into account the fractionalization of the electron into fermionic (Majorana) and charge components \cite{PhysRevX.7.031009}. The electron charge is transported through the interferometer via virtual excitations of the charge degrees of freedom.

For topological superconductors that consist of a semiconductor nanowire proximitized by a bulk superconductor, the time of flight of charge excitations (plasmons) is typically much shorter than the inverse charging energy, and we find that neglecting the effect of charge dynamics on electron tunneling is a good approximation. On the other hand, if the propagation of charge is slowed down, {\em e.g.}, when the superconductor proximitizing the nanowire is (effectively) granular \cite{Science.326.113,PhysRevLett.109.137002,PhysRevLett.109.137003,NatPhys.15.930}, the typical plasmon energy may be less than the charging energy and the coherent electron transfer processes are strongly suppressed. We refer to the superconductor as granular regardless of whether the granularity is intrinsic or the result of intentional engineering. We note that similar tunneling physics is discussed in Refs.\ \onlinecite{PhysRevB.84.144509,PhysRevB.101.125108}; there the topological superconductivity stems from explicitly number-conserving interactions in the quasi-one-dimensional system. By contrast, in our theory we consider topological superconductivity induced by proximity to a three-dimensional superconductor.

\begin{figure}
\begin{center}
\includegraphics[width=0.95\columnwidth]{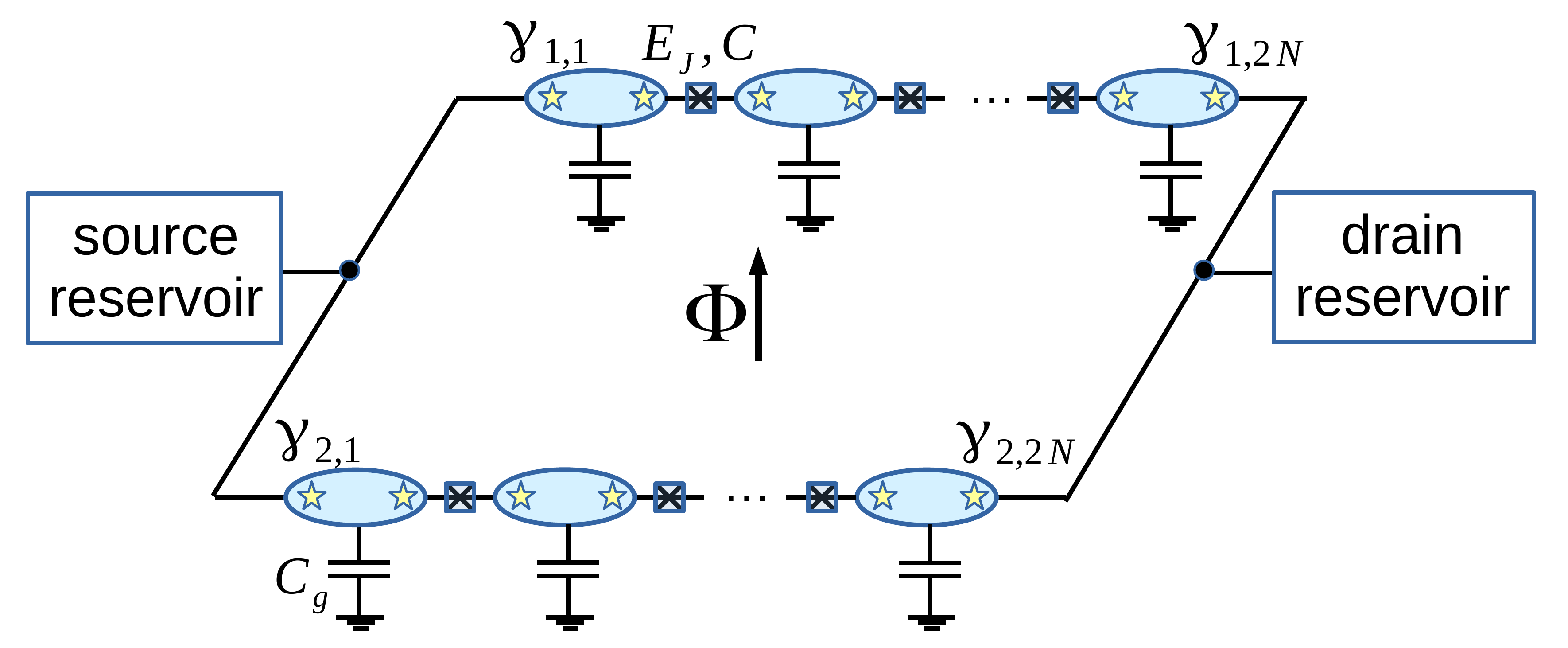}
\end{center}
\caption{Interferometer setup used to measure the coherent tunneling through a Coulomb-blockaded topological superconductor. Each interferometer arm $j=1,2$ is modeled as an array of topological superconductor islands, connected via Josephson junctions. Majorana bound states exist at the ends of every island, but only the Majorana bound states $\gamma_{j,1}$ and $\gamma_{j,2N}$ at the far left and right of the array enter into the low-energy theory. The figure also schematically shows the capacitive coupling between islands and between each island and ground, as well as the flux $\Phi$ through the interference loop.}\label{fig:model}%

\end{figure}

The specific system we consider is an interferometer consisting of source and drain reservoirs connected via two interferometer arms, which are modeled as an array of Josephson junctions connecting islands with topological superconductivity. This is a natural description if the interferometer arm consists of a semiconductor nanowire covered by a granular superconductor. The case of a nanowire with a continuous superconducting cover can be easily obtained as a limiting case of this model. The fermionic low-energy degrees of freedom of each island are Majorana bound states at the two ends of the island, see Fig.\ \ref{fig:model}. Majorana bound states on the two sides of a Josephson junction acquire a finite energy. We assume that the temperature and the applied bias are low enough, that only the Majorana bound states at the far ends of each array of superconducting islands need to be accounted for. 
Hence, for each interferometer arm $j=1,2$, the relevant degrees of freedom are Majorana operators $\gamma_{j,1}$ and $\gamma_{j,2N}$ at the left and right ends of the interferometer arm, see Fig.\ \ref{fig:model}, as well as the charge $n_{j,\alpha}$ of a superconducting island (measured in units of the electron charge $e$) and the conjugate phase variable $\xi_{j,\alpha}$, $[n_{j\alpha},\xi_{j'\alpha'}] = \delta_{jj'} \delta_{\alpha\alpha'}$, where the index $\alpha=1,\ldots,N$ labels the superconducting island. 

Tunneling between the leads and the interferometer arms is described by the Hamiltonian
\begin{align}
  H_{\rm t} =&\, \sum_{j=1}^{2} \sum_{q} \left(t_{j,{\rm L}} c_{q,{\rm L}} \gamma_{j,1} e^{i \xi_{j,1}} 
  \right. \nonumber \\ &\, \left. \ \ \mbox{}
  + t_{j,{\rm R}} c_{q,{\rm R}} \gamma_{j,2N} e^{i \xi_{j,N}} + \mbox{h.c.} \right),
\end{align}
where the operator $e^{i \xi_{j,\alpha}}$ increases $n_{j,\alpha}$ by one and $c_{q,{\rm R}}$ and $c_{q,{\rm L}}$ are the annihilation operators for an electron in the right and left reservoirs at energy $\varepsilon_q$, respectively. To leading order in the tunneling amplitudes, the interference contribution $\delta G$ to the conductance of the interferometer is \footnote{See Supplemental Material at [URL will be inserted by publisher] for the derivation of Eq.\ (\ref{eq:DeltaG}).}
\begin{equation}
  \delta G = \frac{4 \pi e^2}{{\hbar}} p_{1} p_{2} \nu_{\rm L} \nu_{\rm R} 
  \mbox{Re}\, t_{1,{\rm L}}^* {\cal G}_1^{(N)} t_{1,{\rm R}} \,
  t_{2,{\rm L}} {\cal G}_2^{(N)*} t_{2,{\rm R}}^*\,
  e^{i \varphi},
  \label{eq:DeltaG}%
\end{equation}
where $p_j = i \gamma_{j,1} \gamma_{j,2N}$ is the ground-state fermion parity of the $j$th interferometer arm, $\nu_{\rm L}$ and $\nu_{\rm R}$ are the densities of states in the left and right reservoirs, $\varphi/2\pi$ measures the flux through the interferometer in units of $h/e$, and
\begin{equation}
  {\cal G}_j^{(N)} = -i \int_0^{\infty} dt \langle [ e^{-i \xi_{j,1}(t)},e^{i \xi_{j,N}(0)} ] \rangle.
  \label{eq:Gj}
\end{equation}
is the zero-frequency retarded propagator for charge excitations in an array of $N$ superconducting islands. This correlation function also arises in the number-conserving approach of Ref.\ \onlinecite{PhysRevB.101.125108}.

In the case $N=1$ where each arm is modeled as a single island with instantaneous charge redistribution, one has 
\begin{equation}
  {\cal G}_j^{(1)} = - \left( \frac{1}{E_j^+} + \frac{1}{E_j^-} \right),
  \label{eq:GG1}
\end{equation}
where $E_j^{\pm}$ is the energy cost for adding or removing a charge $e$ to the $j$th interferometer arm. In this limit, Eqs.\ (\ref{eq:DeltaG}) and (\ref{eq:Gj}) reproduce the result of Ref.\ \onlinecite{PhysRevLett.104.056402}.

For arbitrary $N$, the charge degrees of freedom of the array are described by the Hamiltonian 
\begin{align}
  H_{{\rm c},j} =&\, \frac{1}{2} e^2 \sum_{\alpha,\alpha'=1}^{N} n_{j,\alpha} C_{\alpha,\alpha'}^{-1} n_{j,\alpha'}
  - e V_{{\rm g},j} \sum_{\alpha=1}^{N_{j}} n_{j,\alpha}
  \nonumber\\ &\, \mbox{}
  - E_{\rm J}^{( 2\pi)} \sum_{\alpha=1}^{N-1} 
   \cos[2(\xi_{j,\alpha} - \xi_{j,\alpha+1})] 
  \nonumber\\ &\, \mbox{}
  - E_{\rm J}^{(4\pi)}\ \sum_{\alpha=1}^{N-1}
   \cos(\xi_{j,\alpha} - \xi_{j,\alpha+1}),
  \label{eq:HWj0}
\end{align}
where, as before, the index $j=1,2$ labels the interferometer arm. Further, $E_{\rm J}^{(2\pi)}$ and $E_{\rm J}^{(4\pi)}$ are $2\pi$- and $4\pi$-periodic Josephson couplings between adjacent islands, $V_{{\rm g},j}$ is a gate voltage, and $C_{\alpha,\alpha'}$ the capacitance matrix,
\begin{equation}
  C_{\alpha\alpha^{\prime}}=C_{\rm g} \delta_{\alpha,\alpha'} + 
  C\left(  2\delta_{\alpha,\alpha'}-\delta_{\alpha,\alpha'+1}-\delta_{\alpha,\alpha'-1}\right),
\end{equation}
where $C_{\rm g}$ is the capacitance between each island and the ground and $C$ is the capacitance between adjacent islands. Note that this model assumes that charge distribution is instantaneous within each island, and that the MBSs within each island do not couple directly. The full phase diagram of this Josephson junction array model in the $C=0$ case is studied in Ref.\ \onlinecite{PhysRevB.101.075419}.
For simplicity, the number of superconducting islands, the capacitances, and the Josephson energies are taken to be identical in the two interferometer arms. Generically one has $C_{\rm g}\ll C$. We consider arrays in the ``transmon regime'', for which the effective Josephson coupling $E_{\rm J} = E_{\rm J}^{(2 \pi)} + (1/4) E_{\rm J}^{(4 \pi)}$ is much larger than the charging energy $e^2/2 C$ associated with the mutual capacitance of neighboring islands. The phase differences $\xi_{j,\alpha} - \xi_{j,\alpha+1}$ are then pinned to the bottom of the cosine potentials, which allows one to disregard the phase slips \cite{PhysRevLett.122.237701} and replace the cosine potentials in Eq.\ (\ref{eq:HWj0}) by a quadratic one. Bringing the Hamiltonian to diagonal form then gives
\begin{align}
  H_{{\rm c},j} \approx&\, E(N_j)
  + \sum_{k=1}^{N-1} \omega_k \left(b_{j,k}^{\dagger} b_{j,k} + \frac{1}{2} \right),
\end{align}
where $N_j = \sum_{\alpha=1}^{N} n_{j,\alpha}$ is the total charge, $E(N_j) = (N_j e)^2/2 N C_{\rm g} - e N_{j} V_{{\rm g},j}$ is the charging energy for a uniformly charged array, 
\begin{equation}
  \omega_k = 2 e \sqrt{\frac{4 E_{\rm J}\sin^2(k \pi/2 N)}{4 C \sin^2(k \pi/2 N)+ C_{\rm g}}} \label{eq:omegak}
\end{equation}
are the plasmon frequencies \cite{PhysRevLett.122.237701}, 
and $b_{j,k}$ are plasmon creation and annihilation operators, which are related to the phase variables as
\begin{align}
  \xi_{j,\alpha} =&\, \Xi_{j} -i \sum_{k=1}^{N-1} \varphi_{\alpha,k} \sqrt{\frac{1}{2 \omega_k C_k}} (b_{j,k} - b^{\dagger}_{j,k}),
  \label{eq:xib}
\end{align}
where $\Xi_j$ is the phase variable conjugate to $N_j$, $C_k = C_{\rm g} + 4 C \sin^2(k \pi/2 N)$, and 
\begin{equation}
  \varphi_{\alpha,k} = \sqrt{\frac{2}{N}} \cos \frac{k \pi (\alpha-1/2)}{N}.
  \label{eq:varphi}
\end{equation}

\begin{figure}
  \includegraphics[width=0.6\columnwidth]{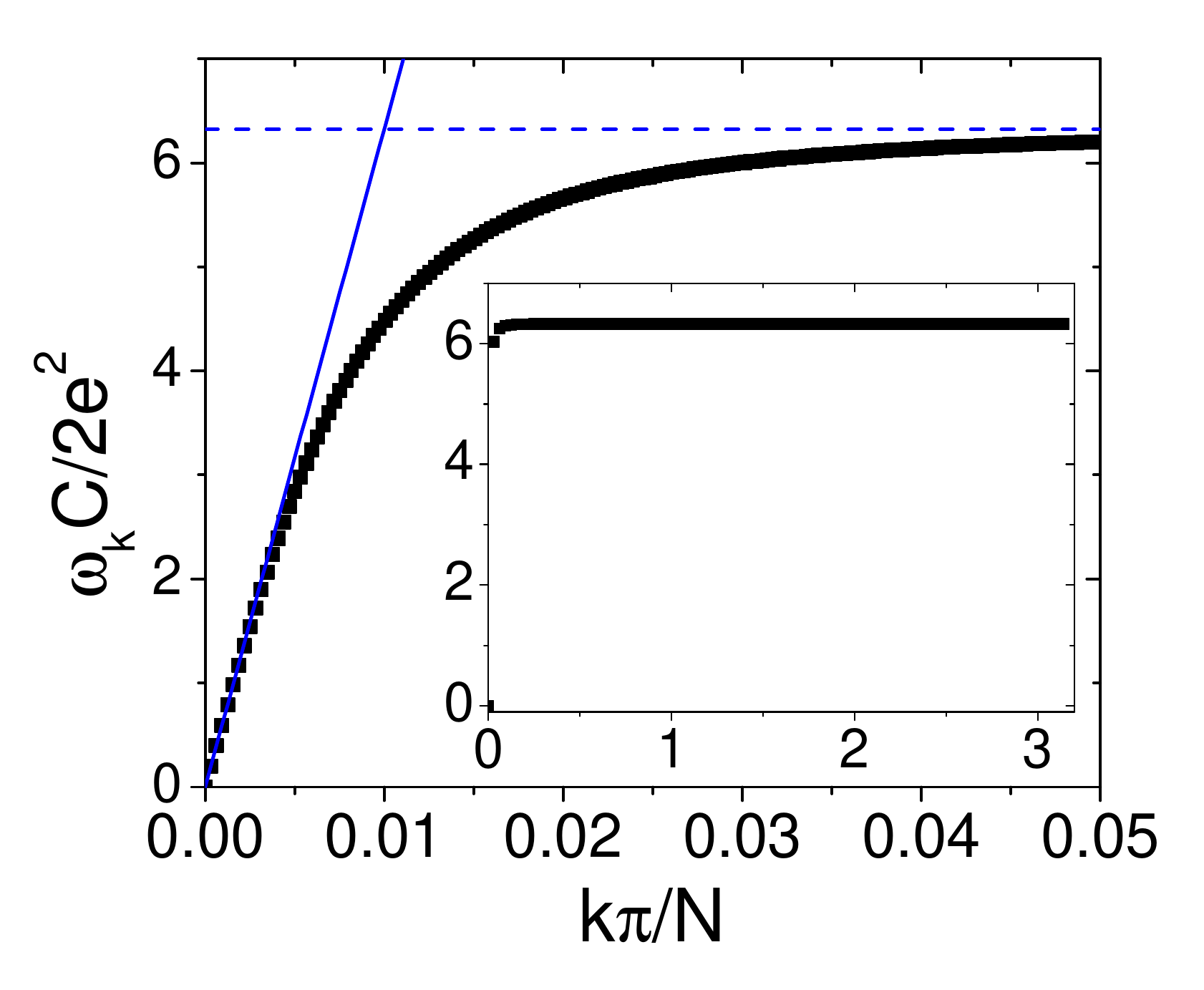}
\caption{Plasmon spectrum of a long ($N=10^{4}$) Josephson junction array in the harmonic approximation, overlaid with the linear (acoustic) approximation at small momenta (solid blue line) and the plasmon energy at $k=N$ (dashed blue line). The parameters chosen are $E_{\rm J} C/e^2=10$ and $C_{\rm g}/C=10^{-4}$. Inset: Plasmon spectrum of a much shorter ($N=100$) Josephson junction array for the same values of $E_{\rm J}$, $C$, and $C_{\rm g}$. The acoustic plasmon branch at small $k$ has almost disappeared.}\label{fig:dispersion_uniform}
\end{figure}

Having brought the Hamitonian $H_j$ to diagonal form, the calculation of the factor ${\cal G}_j$ of Eq.\ (\ref{eq:Gj}) is in principle straightforward. In the cotunneling regime and for temperature $k_{\rm B} T\ll \min(E_j^{\pm})$, one finds
\begin{align}
  {\cal G}_j^{(N)} =&\, - \int_0^{\infty} d\tau 
  \left[ e^{-E_j^+ \tau - \deltaj(\tau)}
  + e^{-E_j^- \tau - \deltaj^*(\tau)} \right],
  \label{eq:Gres1}
\end{align}
with
\begin{align}
  \deltaj(\tau) =&\, \sum_{k=1}^{N-1} \frac{e^{2}(  |\varphi_{1,k}|^{2}+|\varphi_{N,k}|^{2} -  2 \varphi_{1,k}\varphi_{N,k}^* e^{-\omega_{k}\tau})}{4\omega_{k}C_{k}}. 
  \label{eq:Gres2}
\end{align}
Equations (\ref{eq:DeltaG}), (\ref{eq:Gres1}), and (\ref{eq:Gres2}) contain the central results of this work. 
To evaluate the interference contribution $\delta G$ explicitly for the Hamiltonian (\ref{eq:HWj0}), we substitute the explicit expressions for $C_k$ and $\varphi_{j,\alpha}$ and find
\begin{align}
  \deltaj(\tau) =&\, 
  \sum_{k=1}^{N-1} \frac{\omega_k}{16 N E_{\rm J}} [1 - (-1)^k e^{-\omega_k \tau}] \cot^2 \frac{k \pi}{2 N}.
  \label{eq:GtauJJ}
\end{align}

Typically the mutual capacitance $C$ is much larger than the capacitance $C_{\rm g}$ to the ground plane \cite{Science.326.113}. The plasmon dispersion Eq.~(\ref{eq:omegak}) then
interpolates between an acoustic regime\ $\omega_{k}\approx v k \pi/N$ for $k \lesssim k_{\rm c}$, with
\begin{equation}
  v = 2 e \sqrt{\frac{E_{\rm J}}{C_{\rm g}}}, \ \
  k_{\rm c} = \frac{2 N}{\pi} \sqrt{\frac{C_{\rm g}}{4 C + C_{\rm g}}}, \label{eq:v}
\end{equation}
and the constant value $\omega_k \approx \omega_N = 2 e \sqrt{4 E_{\rm J}/(4C+C_{\rm g})}$ when $k \gtrsim k_{\rm c}$, see Fig.~\ref{fig:dispersion_uniform}. 
For short arrays $N \ll \sqrt{C/C_{\rm g}}$, there are no acoustic plasmons and the plasmon frequencies $\omega_k$ are well approximated by $\omega_N$ for all $k$, see Fig.\ \ref{fig:dispersion_uniform}, inset. This gives
\begin{equation}
  \deltaj(\tau) =\, 
  \frac{\omega_N}{16 E_{\rm J}}
  \left[ \frac{N}{3} \left(2 + 
  e^{-\omega_N \tau} \right)
  + 1 -\frac{4}{\pi^2} \right],
  \label{eq:deltatauexp}
\end{equation}
where in addition to taking the limit $N \ll \sqrt{C/C_{\rm g}}$ we expanded in $1/N$, omitting contributions of order $1/N$ and smaller. If both charging energies $E_j^{\pm} \gg \omega_N$, one may approximate $\deltaj(\tau)$ by $\deltaj(0)$ and one finds an exponential suppression of the interference term in the cotunneling current with $N$,
\begin{align}
  {\cal G}_j^{(N)} \approx 
  {\cal G}_j^{(1,{\rm eff})}
  e^{-(\omega_N/16 E_{\rm J}) (N + 1 - 4/\pi^2)},
  \label{eq:Gj1}
\end{align}
where the factor ${\cal G}_j^{(1, {\rm eff})} = -(1/E_{j}^+ + 1/E_{j}^-)$ describes an interferometer arm with a single superconducting island and capacitance $C_{\rm g}^{\rm eff} = N C_{\rm g}$ to the ground plane \cite{PhysRevLett.104.056402}, see Eq.\ (\ref{eq:GG1}).
(If the condition $E_{j}^{\pm} \gg \omega_N$ is not met, there is still an exponential suppression with $N$, but with a numerically different exponent.) 
To understand the exponential dependence on $N$, notice that the phase differences between adjacent islands $\xi_{j,\alpha}-\xi_{j,\alpha+1}$ ($\alpha=1,\ldots,N-1$) are independent variables in the limit $C_g\to 0$, as follows from Eq.~(\ref{eq:xib}); ${\cal G}_j^{(N)}$ thus factorizes into identical contributions from individual Josephson junctions.

\begin{figure}[h]
\includegraphics[width=1\columnwidth]{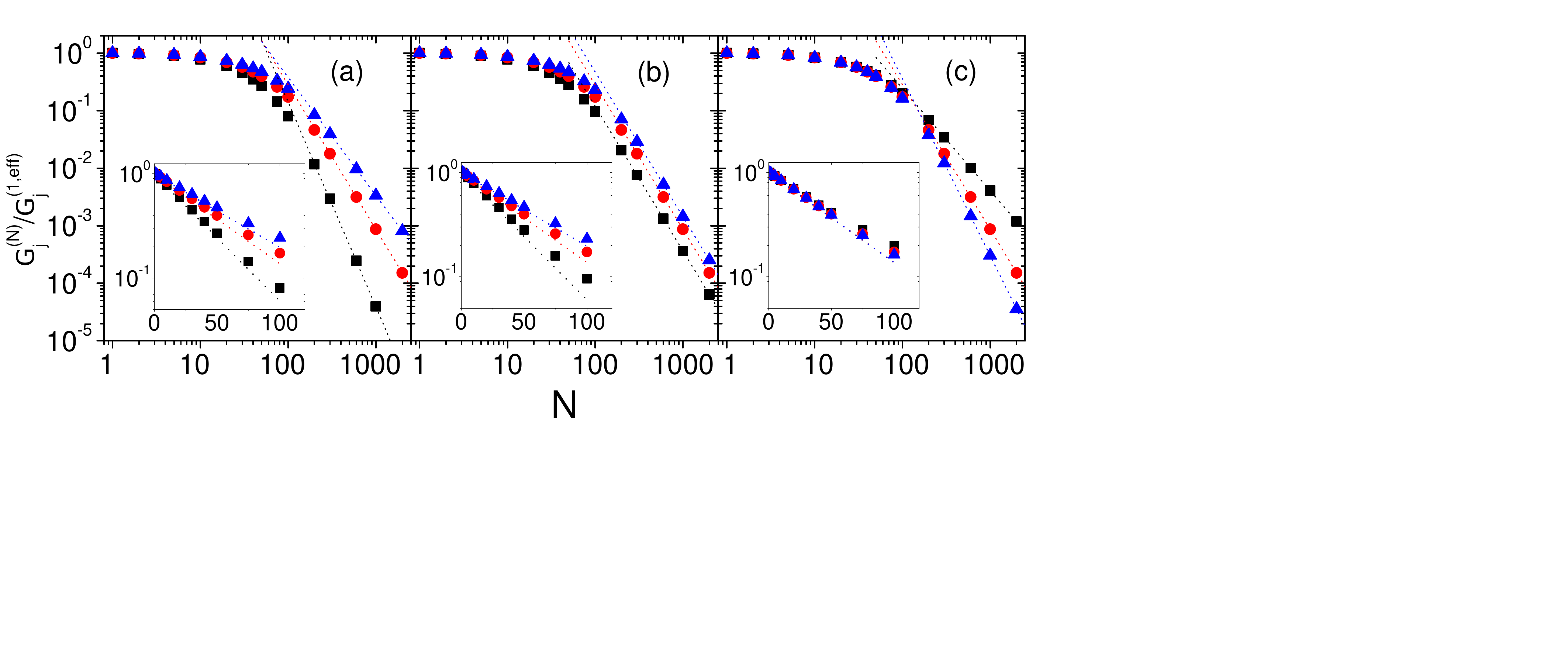}\caption{Normalized coherent amplitude ${\cal G}_j^{(N)}/{\cal G}_j^{(1,{\rm eff})}$ as a function of the number of superconducting islands $N$. The insets focus on the exponential $N$ dependence in short systems. The data points show the amplitude ratio ${\cal G}_j^{(N)}/{\cal G}_j^{(1,{\rm eff})}$ for $E_{\rm J} C/e^2 = 5$, $10$, and $15$ and $C_{\rm g}/C = 10^{-4}$ (bottom to top data sets, left), for $E_{\rm J} C/e^2 = 5$, $10$, and $15$ and $E_{\rm J} C_{\rm g}/e^2 = 1 \times 10^{-3}$ (bottom to top, center) and $C/C_{\rm g} = 5 \times 10^3$, $10^4$, $1.5 \times 10^4$ and $E_{\rm J} C/e^2 = 10$ (top to bottom, right). The dotted curves follow from the asymptotic expressions (\ref{eq:deltatauexp}) and (\ref{eq:deltataupow}).} \label{fig:current_uniform}%
\end{figure}

%
For long arrays, $N \gg \sqrt{C/C_{\rm g}}$, the summation (\ref{eq:GtauJJ}) is dominated by the acoustic branch $\omega_k \approx v k \pi/N$ for $k \ll k_{\rm c}$, see Eq.\ (\ref{eq:v}). In this regime, it is instructive to express $\delta_j(\tau)$ in terms of the parameters $v$ and $k_{\rm c}$,
\begin{equation}
  \delta_j(\tau) =\, \frac{v}{4 \pi E_{\rm J}} \left\{\ln\left[2 k_{\rm c} (1 + e^{-\pi v \tau/N})\right] - f(C_{\rm g}/4C) \right\},
  \label{eq:deltataupow}
\end{equation}
where $f(x) = \sqrt{x}\, \mbox{arccot}\sqrt{x} -  c_{\gamma}$ and $c_{\gamma} \approx 0.577$ is the Euler-Mascheroni constant. 
As before, we may approximate $\deltaj(\tau)$ by $\deltaj(0)$ if both charging energies $E_j^{\pm} \gg \pi v/N$. This gives a power-law suppression with $N$ (recall $k_c\propto N$),
\begin{equation}
  {\cal G}_j^{(N)} \approx {\cal G}_j^{(1,{\rm eff})}
  (4 k_{\rm c})^{-\beta}
  e^{\beta f(C_{\rm g}/4C)},\ \ \beta = \frac{v}{4 \pi E_{\rm J}}.
  \label{eq:Napow}
\end{equation}
Note that the exponent $\beta$ is independent of the capacitance $C$ between adjacent islands, because $C$ does not enter into the low-energy degrees of freedom. (Again, if the condition $E_j^{\pm} \gg \pi v/N$ is not met, there is still a power-law suppression with $N$ but with a different numerical prefactor.) In Fig.\ \ref{fig:current_uniform} we show ${\cal G}_j$ as a function of $N$ for both sides of the crossover at $N \sim \sqrt{C/C_{\rm g}}$ and compare with the predictions of the asymptotic expressions (\ref{eq:deltatauexp}) and (\ref{eq:deltataupow}).

The power law of Eq.~(\ref{eq:Napow}) can also be obtained from a continuum model in which the charge degrees of freedom are described as a transmission line with capacitance $c$ and inductance $\ell$ per unit length \cite{*[{See, for example, }][{}] devoret2014quantum}. This requires one to identify $v = 1/\sqrt{\ell c}$ and $E_{\rm J} = 1/4 e^2\ell$ \cite{PhysRevLett.122.237701}, so that the exponent $\beta = e^2/\pi \sqrt{\ell/c}$. In this continuum description, $k_{\rm c}$ is the ultraviolet cutoff of the theory, which signals the breakdown of the one-dimensional linear plasmon dispersion. 

If the topological superconductor consists of a semiconducting nanowire covered by well-separated superconducting grains, the Josephson coupling $E_{\rm J} \sim g_{\rm c} \Delta$, where $\Delta$ is the magnitude of the superconducting gap and $g_{\rm c} \gtrsim 1$ is the dimensionless conductance of a wire segment connecting two superconducting islands. Taking a typical charging energy $e^2/2 C \sim E_{\rm J}/10$ \cite{PhysRevLett.122.237701}, we find that the suppression of the interference contribution to the conductance can be appreciable even for moderate values of $N$, as shown in Fig.\ \ref{fig:current_uniform}. On the other hand, if the topological superconductor consists of a semiconductor nanowire covered by a single superconductor, a continuum description in terms of a capacitance $c$ and inductance $\ell$ per unit length is more applicable. Typically, $e^{-2} \sqrt{\ell/c}$ is between $50$ and $300$ Ohms, which places the exponent $\beta$ between $10^{-3}$ and $10^{-2}$. Hence, for a covering with a continuous superconductor the suppression of the interference term is usually weak, independent of the value of the ultraviolet cutoff $k_{\rm c}$, and to a very good approximation, the coherent cotunneling process is given by the result Eq.~(\ref{eq:GG1}) for a single superconducting island. A posteriori, this justifies the assumption of instantaneous charge redistribution within each island which is implicit in the model of Eq.\ (\ref{eq:HWj0}).

The physics of the absence of suppression for the continuum description is analogous to that of the environmental Coulomb blockade \cite{ZhEkspTeorFiz.95.975,*SovPhysJETP.68.561,*PismaZhEkspTeorFiz.49.105,*JETPLett.49.126,PhysRevLett.64.1824,PhysRevLett.64.3183}: the Coulomb blockade is suppressed by the discharging of the tunnel junction when the $RC$ time for charge displacements is much smaller than the Heisenberg uncertainty time $\hbar C/e^2$. Similarly, for a system of length $L$, the typical time scale for charge redistribution $L/v$ is much shorter than $\hbar C/e^2$, or in other words, the plasmon quantization energy is much larger than the charging energy. Up to a factor, the resulting exponent $\beta$ is the same as the exponent found for the power law suppression of the differential conductance in the environmental Coulomb blockade. 

We close by remarking that our results can be easily generalized to the case where one or more tunnel junctions (weak links) exist in the arms, as occurs in various Majorana network models and stabilizer measurements in corresponding implementations of topological quantum error correction codes \cite{PhysRevX.5.041038,PhysRevB.94.174514,PhysRevX.7.031048}. To calculate the lowest-order interference contribution to the cotunneling current, one takes all weakly coupled segments in the interference loop, and multiplies their tunneling amplitudes, fermion parities, and suppression factors ${\cal G}_{j}$.



In summary, we have studied the coherence of cotunneling of single electrons
through Majorana wires. In contrast to previous studies we have included
the charge degrees of freedom in addition to the fermion component. For
semiconductor nanowires proximitized by bulk superconductors, the typical plasmon energy is large
compared to the charging energy, and we show the cotunneling transmission
amplitude is to a very good approximation given by the fermion-only expression
in Ref.~\onlinecite{PhysRevLett.104.056402}. On the other hand, for nanowires proximitized by superconducting islands which form Josephson junction arrays operating in the transmon
regime, the typical plasmon energy is usually much smaller than the charging energy,
and as a result we find the coherent cotunneling to be significantly suppressed by charge dynamics.

We acknowledge support by CRC-TR 183 (project C03 and Mercator professorship) of Deutsche Forschungsgemeinschaft (ZS, PWB, KF, FvO), the Danish National Research Foundation (KF), Independent Research Fund Denmark $|$ Natural Sciences (KF), the DOE contract DE-FG02-08ER46482 (LIG), and QuantERA project {\em Topoquant} (FvO).

\appendix

\begin{widetext}

\section{Derivation of Eq.~(2) in the main text}

In this Appendix, we derive the interference contribution to the conductance,
Eq.~(2) in the main text. Following the closing remarks of the main text, we
analyze the slightly more general model with $N_{j}-1$ weak links in arm $j$%
\ ($j=1,2$). The tunneling Hamiltonian takes the form
\begin{equation}
H_{\text{t}}=\sum_{j=1}^{2}\left[ \sum_{q}\left( t_{j,\text{L}}c_{q,\text{L}%
}\gamma _{j,1}e^{i\xi _{j,1,\text{L}}}+t_{j,\text{R}}c_{q,\text{R}}\gamma
_{j,2N_{j}}e^{i\xi _{j,N_{j},\text{R}}}\right)
-\sum_{l=1}^{N_{j}-1}it_{j,l}\gamma _{j,2l+1}e^{-i\xi _{j,l+1,\text{L}%
}}\gamma _{j,2l}e^{i\xi _{j,l,\text{R}}}\right] +\text{h.c.}
\end{equation}%
where the $l$th weak link in arm $j$ has a hopping amplitude of $t_{j,l}$.
We have relabeled the Majorana zero modes (MZMs) so that $\gamma _{j,2l-1}$ and $\gamma _{j,2l}$
now represent the MZMs at the left and right ends of the $l$th weakly
coupled segment in arm $j$, and $e^{i\xi _{j,l,\text{L/R}}}$ are the
corresponding charge creation operators. The remaining unlabeled MZMs do not
appear in the low-energy theory. We also absorb 
the phase factor $e^{i\varphi }$ due to the flux through the interferometer
in the hopping amplitudes.

The current operator for the current in the right lead has the form
\begin{align}
  I_{\text{R}}=&\,
  i\frac{e}{\hbar }\left[ H_{\text{t}},\sum_{q}c_{q,\text{R}%
}^{\dag }c_{q,\text{R}}\right] \nonumber \\ =&
  i\frac{e}{\hbar }\sum_{j=1}^{2}\sum_{q}%
\left( t_{j,\text{R}}c_{q,\text{R}}\gamma _{j,2N_{j}}e^{i\xi _{j,N_{j},\text{%
R}}}-t_{j,\text{R}}^{\ast }\gamma _{j,2N_{j}}e^{-i\xi _{j,N_{j},\text{R}%
}}c_{q,\text{R}}^{\dag }\right) \text{,}
\end{align}%
so it is natural to consider the contour-ordered Green function

\begin{equation}
G_{I_{\text{R}}}\left( t,t^{\prime }\right) =i\frac{e}{\hbar }\left\langle
T_{c}\sum_{j=1}^{2}\sum_{q}\left[ t_{j,\text{R}}c_{q,\text{R}}\left(
t\right) \gamma _{j,2N_{j}}\left( t^{\prime }\right) e^{i\xi _{j,N_{j},\text{%
R}}(t^{\prime })}-t_{j,\text{R}}^{\ast }\gamma _{j,2N_{j}}\left( t^{\prime
}\right) e^{-i\xi _{j,N_{j},\text{R}}(t^{\prime })}c_{q,\text{R}}^{\dag
}\left( t\right) \right] \right\rangle _{\text{H}}\text{.}
\end{equation}%
Here $T_{c}$ is the time-ordering operator on the Keldysh contour.
The lowest-order contributions
to the interference term in the cotunneling current correspond to the
following terms in the Dyson series of $G_{I_{\text{R}}}$,
\begin{align}
\delta G_{I_{\text{R}}}& =\frac{e}{\hbar }\sum_{qq^{\prime }}\left\{ t_{1,%
\text{R}}t_{2,\text{R}}^{\ast }t_{1,\text{L}}^{\ast }t_{2,\text{L}}t_{1,%
\text{tot}}t_{2,\text{tot}}^{\ast }\left[ G_{q,\text{R}}G_{2,\text{RL}%
}G_{q^{\prime },\text{L}}G_{1,\text{LR}}-\left( G_{2,\text{RL}}G_{q^{\prime
},\text{L}}G_{1,\text{LR}}G_{q,\text{R}}\right) ^{\text{T}}\right] \right.  
\notag \\
& \left. +t_{1,\text{R}}^{\ast }t_{2,\text{R}}t_{1,\text{L}}t_{2,\text{L}%
}^{\ast }t_{1,\text{tot}}^{\ast }t_{2,\text{tot}}\left[ G_{q,\text{R}}G_{1,%
\text{RL}}G_{q^{\prime },\text{L}}G_{2,\text{LR}}-\left( G_{1,\text{RL}%
}G_{q^{\prime },\text{L}}G_{2,\text{LR}}G_{q,\text{R}}\right) ^{\text{T}}%
\right] \right\} \text{.}
\end{align}%
We have defined $t_{j,\text{tot}}\equiv \prod_{l=1}^{N_{j}-1}\left(
it_{j,l}\right) $, understanding that $t_{j,\text{tot}}=1$ if $N_{j}=1$. All
Green functions have an implicit matrix structure in their time arguments
(on which the transposition operates), and the matrix multiplication
operation is identified as the Keldysh contour time convolution, $\left[
G_{1}G_{2}\right] \left( t,t^{\prime }\right) \equiv \int_{c}dt^{\prime
\prime }G_{1}\left( t,t^{\prime \prime }\right) G_{2}\left( t^{\prime \prime
},t^{\prime }\right) $. The lead Green functions are given by

\begin{equation}
G_{q,\alpha }\left( t,t^{\prime }\right) =-i\left\langle T_{c}c_{q,\alpha
}\left( t\right) c_{q,\alpha }^{\dag }\left( t^{\prime }\right)
\right\rangle \text{, }\alpha =\text{L, R,}
\end{equation}%
and the charge sector Green functions

\begin{equation}
G_{j,\text{RL}}\equiv G_{j,N_{j},\text{RL}}G_{j,N_{j}-1,\text{RL}}\cdots
G_{j,1,\text{RL}}\text{,}
\end{equation}

\begin{equation}
G_{j,\text{LR}}\equiv G_{j,1,\text{LR}}G_{j,2,\text{LR}}\cdots G_{j,N_{j},%
\text{LR}}\text{,}
\end{equation}%
where

\begin{equation}
G_{j,l,\text{LR}}\left( t,t^{\prime }\right) =-i\left\langle T_{c}\gamma
_{j,2l-1}\left( t\right) e^{-i\xi _{j,l,\text{L}}\left( t\right) }\gamma
_{j,2l}\left( t^{\prime }\right) e^{i\xi _{j,l,\text{R}}(t^{\prime
})}\right\rangle \text{,}
\end{equation}

\begin{equation}
G_{j,l,\text{RL}}\left( t,t^{\prime }\right) =-i\left\langle T_{c}\gamma
_{j,2l}\left( t\right) e^{-i\xi _{j,l,\text{R}}\left( t\right) }\gamma
_{j,2l-1}\left( t^{\prime }\right) e^{i\xi _{j,l,\text{L}}(t^{\prime
})}\right\rangle \text{.}
\end{equation}%
The expectation value of the current is the lesser Green function $I_{\text{R%
}}\left( t\right) =G_{I_{\text{R}}}^{<}\left( t,t\right) $. Analytic
continuation yields in the frequency domain

\begin{align}
\delta I_{\text{R}}& =\frac{e}{\hbar }\int \frac{d\omega }{2\pi }%
\sum_{qq^{\prime }}\left[ t_{1,\text{R}}t_{2,\text{R}}^{\ast }t_{1,\text{L}%
}^{\ast }t_{2,\text{L}}t_{1,\text{tot}}t_{2,\text{tot}}^{\ast }\left( G_{q,%
\text{R}}^{\text{R}}G_{2,\text{RL}}^{\text{R}}G_{q^{\prime },\text{L}%
}^{<}G_{1,\text{LR}}^{\text{A}}\right. \right.   \notag \\
& \left. \left. +G_{q,\text{R}}^{<}G_{2,\text{RL}}^{\text{A}}G_{q^{\prime },%
\text{L}}^{\text{A}}G_{1,\text{LR}}^{\text{A}}-G_{2,\text{RL}}^{\text{R}%
}G_{q^{\prime },\text{L}}^{\text{R}}G_{1,\text{LR}}^{\text{R}}G_{q,\text{R}%
}^{>}-G_{2,\text{RL}}^{\text{R}}G_{q^{\prime },\text{L}}^{>}G_{1,\text{LR}}^{%
\text{A}}G_{q,\text{R}}^{\text{A}}\right) +\text{c.c.}\right] \text{,}
\end{align}%
where we have suppressed the frequency argument $\omega $ (identical for all
Green functions), and discarded the charge sector lesser and greater Green
functions on the grounds that all charge excitations are virtual in the
cotunneling regime. If the bias voltage $V$ and the temperature $T$
satisfies $\left\vert eV\right\vert $, $T\ll E_{j}^{\pm }$, it is
permissible to ignore the $\omega $ dependence of the charge sector retarded
and advanced Green functions and approximate

\begin{equation}
G_{j,l,\text{LR}}^{\text{R}}\left( \omega \right) \approx -ip_{j,l}\mathcal{G%
}_{j,l}\text{, }G_{j,l,\text{RL}}^{\text{R}}\left( \omega \right) \approx
ip_{j,l}\mathcal{G}_{j,l}^{\ast }\text{,}
\end{equation}%
where $p_{j,l}\equiv i\gamma _{j,2l-1}\gamma _{j,2l}$ is the ground-state fermion parity
of the $l$th weakly coupled segment in arm $j$, and

\begin{equation}
\mathcal{G}_{j,l}\equiv -i\int_{0}^{\infty }dt\left\langle \left[ e^{-i\xi
_{j,l,\text{L}}\left( t\right) },e^{i\xi _{j,l,\text{R}}\left( 0\right) }%
\right] \right\rangle 
\end{equation}%
generalizes the definition Eq.~(3) in the main text. The remaining integral and
summations are straightforward, and we eventually find the interference
contribution to the cotunneling current

\begin{equation}
\delta I_{\text{R}}=V\frac{4\pi e^{2}}{\hbar }\nu _{\text{L}}\nu _{\text{R}}%
\operatorname{Re}\left[ t_{1,\text{R}}t_{2,\text{R}}^{\ast }t_{1,\text{L}}^{\ast
}t_{2,\text{L}}\prod_{l=1}^{N_{1}-1}t_{1,l}\prod_{l=1}^{N_{2}-1}t_{2,l}^{%
\ast }\prod_{l=1}^{N_{1}}\left( p_{1,l}\mathcal{G}_{1,l}\right)
\prod_{l=1}^{N_{2}}\left( p_{2,l}\mathcal{G}_{2,l}^{\ast }\right) \right] 
\text{.}
\end{equation}%
In the special case $N_{1}=N_{2}=1$, this immediately reproduces
Eq.~(2) in the main text.

\end{widetext}
\bibliography{teleport}

\begin{thebibliography}{32}%
\makeatletter
\providecommand \@ifxundefined [1]{%
 \@ifx{#1\undefined}
}%
\providecommand \@ifnum [1]{%
 \ifnum #1\expandafter \@firstoftwo
 \else \expandafter \@secondoftwo
 \fi
}%
\providecommand \@ifx [1]{%
 \ifx #1\expandafter \@firstoftwo
 \else \expandafter \@secondoftwo
 \fi
}%
\providecommand \natexlab [1]{#1}%
\providecommand \enquote  [1]{``#1''}%
\providecommand \bibnamefont  [1]{#1}%
\providecommand \bibfnamefont [1]{#1}%
\providecommand \citenamefont [1]{#1}%
\providecommand \href@noop [0]{\@secondoftwo}%
\providecommand \href [0]{\begingroup \@sanitize@url \@href}%
\providecommand \@href[1]{\@@startlink{#1}\@@href}%
\providecommand \@@href[1]{\endgroup#1\@@endlink}%
\providecommand \@sanitize@url [0]{\catcode `\\12\catcode `\$12\catcode
  `\&12\catcode `\#12\catcode `\^12\catcode `\_12\catcode `\%12\relax}%
\providecommand \@@startlink[1]{}%
\providecommand \@@endlink[0]{}%
\providecommand \url  [0]{\begingroup\@sanitize@url \@url }%
\providecommand \@url [1]{\endgroup\@href {#1}{\urlprefix }}%
\providecommand \urlprefix  [0]{URL }%
\providecommand \Eprint [0]{\href }%
\providecommand \doibase [0]{http://dx.doi.org/}%
\providecommand \selectlanguage [0]{\@gobble}%
\providecommand \bibinfo  [0]{\@secondoftwo}%
\providecommand \bibfield  [0]{\@secondoftwo}%
\providecommand \translation [1]{[#1]}%
\providecommand \BibitemOpen [0]{}%
\providecommand \bibitemStop [0]{}%
\providecommand \bibitemNoStop [0]{.\EOS\space}%
\providecommand \EOS [0]{\spacefactor3000\relax}%
\providecommand \BibitemShut  [1]{\csname bibitem#1\endcsname}%
\let\auto@bib@innerbib\@empty
\bibitem [{\citenamefont {Nayak}\ \emph {et~al.}(2008)\citenamefont {Nayak},
  \citenamefont {Simon}, \citenamefont {Stern}, \citenamefont {Freedman},\ and\
  \citenamefont {Das~Sarma}}]{RevModPhys.80.1083}%
  \BibitemOpen
  \bibfield  {author} {\bibinfo {author} {\bibfnamefont {C.}~\bibnamefont
  {Nayak}}, \bibinfo {author} {\bibfnamefont {S.~H.}\ \bibnamefont {Simon}},
  \bibinfo {author} {\bibfnamefont {A.}~\bibnamefont {Stern}}, \bibinfo
  {author} {\bibfnamefont {M.}~\bibnamefont {Freedman}}, \ and\ \bibinfo
  {author} {\bibfnamefont {S.}~\bibnamefont {Das~Sarma}},\ }\href {\doibase
  10.1103/RevModPhys.80.1083} {\bibfield  {journal} {\bibinfo  {journal} {Rev.
  Mod. Phys.}\ }\textbf {\bibinfo {volume} {80}},\ \bibinfo {pages} {1083}
  (\bibinfo {year} {2008})}\BibitemShut {NoStop}%
\bibitem [{\citenamefont {Hyart}\ \emph {et~al.}(2013)\citenamefont {Hyart},
  \citenamefont {van Heck}, \citenamefont {Fulga}, \citenamefont {Burrello},
  \citenamefont {Akhmerov},\ and\ \citenamefont
  {Beenakker}}]{PhysRevB.88.035121}%
  \BibitemOpen
  \bibfield  {author} {\bibinfo {author} {\bibfnamefont {T.}~\bibnamefont
  {Hyart}}, \bibinfo {author} {\bibfnamefont {B.}~\bibnamefont {van Heck}},
  \bibinfo {author} {\bibfnamefont {I.~C.}\ \bibnamefont {Fulga}}, \bibinfo
  {author} {\bibfnamefont {M.}~\bibnamefont {Burrello}}, \bibinfo {author}
  {\bibfnamefont {A.~R.}\ \bibnamefont {Akhmerov}}, \ and\ \bibinfo {author}
  {\bibfnamefont {C.~W.~J.}\ \bibnamefont {Beenakker}},\ }\href {\doibase
  10.1103/PhysRevB.88.035121} {\bibfield  {journal} {\bibinfo  {journal} {Phys.
  Rev. B}\ }\textbf {\bibinfo {volume} {88}},\ \bibinfo {pages} {035121}
  (\bibinfo {year} {2013})}\BibitemShut {NoStop}%
\bibitem [{\citenamefont {Vijay}\ \emph {et~al.}(2015)\citenamefont {Vijay},
  \citenamefont {Hsieh},\ and\ \citenamefont {Fu}}]{PhysRevX.5.041038}%
  \BibitemOpen
  \bibfield  {author} {\bibinfo {author} {\bibfnamefont {S.}~\bibnamefont
  {Vijay}}, \bibinfo {author} {\bibfnamefont {T.~H.}\ \bibnamefont {Hsieh}}, \
  and\ \bibinfo {author} {\bibfnamefont {L.}~\bibnamefont {Fu}},\ }\href
  {\doibase 10.1103/PhysRevX.5.041038} {\bibfield  {journal} {\bibinfo
  {journal} {Phys. Rev. X}\ }\textbf {\bibinfo {volume} {5}},\ \bibinfo {pages}
  {041038} (\bibinfo {year} {2015})}\BibitemShut {NoStop}%
\bibitem [{\citenamefont {Plugge}\ \emph {et~al.}(2016)\citenamefont {Plugge},
  \citenamefont {Landau}, \citenamefont {Sela}, \citenamefont {Altland},
  \citenamefont {Flensberg},\ and\ \citenamefont {Egger}}]{PhysRevB.94.174514}%
  \BibitemOpen
  \bibfield  {author} {\bibinfo {author} {\bibfnamefont {S.}~\bibnamefont
  {Plugge}}, \bibinfo {author} {\bibfnamefont {L.~A.}\ \bibnamefont {Landau}},
  \bibinfo {author} {\bibfnamefont {E.}~\bibnamefont {Sela}}, \bibinfo {author}
  {\bibfnamefont {A.}~\bibnamefont {Altland}}, \bibinfo {author} {\bibfnamefont
  {K.}~\bibnamefont {Flensberg}}, \ and\ \bibinfo {author} {\bibfnamefont
  {R.}~\bibnamefont {Egger}},\ }\href {\doibase 10.1103/PhysRevB.94.174514}
  {\bibfield  {journal} {\bibinfo  {journal} {Phys. Rev. B}\ }\textbf {\bibinfo
  {volume} {94}},\ \bibinfo {pages} {174514} (\bibinfo {year}
  {2016})}\BibitemShut {NoStop}%
\bibitem [{\citenamefont {Karzig}\ \emph {et~al.}(2017)\citenamefont {Karzig},
  \citenamefont {Knapp}, \citenamefont {Lutchyn}, \citenamefont {Bonderson},
  \citenamefont {Hastings}, \citenamefont {Nayak}, \citenamefont {Alicea},
  \citenamefont {Flensberg}, \citenamefont {Plugge}, \citenamefont {Oreg},
  \citenamefont {Marcus},\ and\ \citenamefont {Freedman}}]{PhysRevB.95.235305}%
  \BibitemOpen
  \bibfield  {author} {\bibinfo {author} {\bibfnamefont {T.}~\bibnamefont
  {Karzig}}, \bibinfo {author} {\bibfnamefont {C.}~\bibnamefont {Knapp}},
  \bibinfo {author} {\bibfnamefont {R.~M.}\ \bibnamefont {Lutchyn}}, \bibinfo
  {author} {\bibfnamefont {P.}~\bibnamefont {Bonderson}}, \bibinfo {author}
  {\bibfnamefont {M.~B.}\ \bibnamefont {Hastings}}, \bibinfo {author}
  {\bibfnamefont {C.}~\bibnamefont {Nayak}}, \bibinfo {author} {\bibfnamefont
  {J.}~\bibnamefont {Alicea}}, \bibinfo {author} {\bibfnamefont
  {K.}~\bibnamefont {Flensberg}}, \bibinfo {author} {\bibfnamefont
  {S.}~\bibnamefont {Plugge}}, \bibinfo {author} {\bibfnamefont
  {Y.}~\bibnamefont {Oreg}}, \bibinfo {author} {\bibfnamefont {C.~M.}\
  \bibnamefont {Marcus}}, \ and\ \bibinfo {author} {\bibfnamefont {M.~H.}\
  \bibnamefont {Freedman}},\ }\href {\doibase 10.1103/PhysRevB.95.235305}
  {\bibfield  {journal} {\bibinfo  {journal} {Phys. Rev. B}\ }\textbf {\bibinfo
  {volume} {95}},\ \bibinfo {pages} {235305} (\bibinfo {year}
  {2017})}\BibitemShut {NoStop}%
\bibitem [{\citenamefont {Vijay}\ and\ \citenamefont
  {Fu}(2016)}]{PhysRevB.94.235446}%
  \BibitemOpen
  \bibfield  {author} {\bibinfo {author} {\bibfnamefont {S.}~\bibnamefont
  {Vijay}}\ and\ \bibinfo {author} {\bibfnamefont {L.}~\bibnamefont {Fu}},\
  }\href {\doibase 10.1103/PhysRevB.94.235446} {\bibfield  {journal} {\bibinfo
  {journal} {Phys. Rev. B}\ }\textbf {\bibinfo {volume} {94}},\ \bibinfo
  {pages} {235446} (\bibinfo {year} {2016})}\BibitemShut {NoStop}%
\bibitem [{\citenamefont {Plugge}\ \emph {et~al.}(2017)\citenamefont {Plugge},
  \citenamefont {Rasmussen}, \citenamefont {Egger},\ and\ \citenamefont
  {Flensberg}}]{NewJPhys.19.012001}%
  \BibitemOpen
  \bibfield  {author} {\bibinfo {author} {\bibfnamefont {S.}~\bibnamefont
  {Plugge}}, \bibinfo {author} {\bibfnamefont {A.}~\bibnamefont {Rasmussen}},
  \bibinfo {author} {\bibfnamefont {R.}~\bibnamefont {Egger}}, \ and\ \bibinfo
  {author} {\bibfnamefont {K.}~\bibnamefont {Flensberg}},\ }\href {\doibase
  10.1088/1367-2630/aa54e1} {\bibfield  {journal} {\bibinfo  {journal} {New J.
  Phys.}\ }\textbf {\bibinfo {volume} {19}},\ \bibinfo {pages} {012001}
  (\bibinfo {year} {2017})}\BibitemShut {NoStop}%
\bibitem [{\citenamefont {Litinski}\ \emph {et~al.}(2017)\citenamefont
  {Litinski}, \citenamefont {Kesselring}, \citenamefont {Eisert},\ and\
  \citenamefont {von Oppen}}]{PhysRevX.7.031048}%
  \BibitemOpen
  \bibfield  {author} {\bibinfo {author} {\bibfnamefont {D.}~\bibnamefont
  {Litinski}}, \bibinfo {author} {\bibfnamefont {M.~S.}\ \bibnamefont
  {Kesselring}}, \bibinfo {author} {\bibfnamefont {J.}~\bibnamefont {Eisert}},
  \ and\ \bibinfo {author} {\bibfnamefont {F.}~\bibnamefont {von Oppen}},\
  }\href {\doibase 10.1103/PhysRevX.7.031048} {\bibfield  {journal} {\bibinfo
  {journal} {Phys. Rev. X}\ }\textbf {\bibinfo {volume} {7}},\ \bibinfo {pages}
  {031048} (\bibinfo {year} {2017})}\BibitemShut {NoStop}%
\bibitem [{\citenamefont {Litinski}\ and\ \citenamefont {von
  Oppen}(2018)}]{PhysRevB.97.205404}%
  \BibitemOpen
  \bibfield  {author} {\bibinfo {author} {\bibfnamefont {D.}~\bibnamefont
  {Litinski}}\ and\ \bibinfo {author} {\bibfnamefont {F.}~\bibnamefont {von
  Oppen}},\ }\href {\doibase 10.1103/PhysRevB.97.205404} {\bibfield  {journal}
  {\bibinfo  {journal} {Phys. Rev. B}\ }\textbf {\bibinfo {volume} {97}},\
  \bibinfo {pages} {205404} (\bibinfo {year} {2018})}\BibitemShut {NoStop}%
\bibitem [{\citenamefont {Fu}(2010)}]{PhysRevLett.104.056402}%
  \BibitemOpen
  \bibfield  {author} {\bibinfo {author} {\bibfnamefont {L.}~\bibnamefont
  {Fu}},\ }\href {\doibase 10.1103/PhysRevLett.104.056402} {\bibfield
  {journal} {\bibinfo  {journal} {Phys. Rev. Lett.}\ }\textbf {\bibinfo
  {volume} {104}},\ \bibinfo {pages} {056402} (\bibinfo {year}
  {2010})}\BibitemShut {NoStop}%
\bibitem [{\citenamefont {Lutchyn}\ \emph {et~al.}(2018)\citenamefont
  {Lutchyn}, \citenamefont {Bakkers}, \citenamefont {Kouwenhoven},
  \citenamefont {Krogstrup}, \citenamefont {Marcus},\ and\ \citenamefont
  {Oreg}}]{NatRevMater.3.52}%
  \BibitemOpen
  \bibfield  {author} {\bibinfo {author} {\bibfnamefont {R.~M.}\ \bibnamefont
  {Lutchyn}}, \bibinfo {author} {\bibfnamefont {E.~P. A.~M.}\ \bibnamefont
  {Bakkers}}, \bibinfo {author} {\bibfnamefont {L.~P.}\ \bibnamefont
  {Kouwenhoven}}, \bibinfo {author} {\bibfnamefont {P.}~\bibnamefont
  {Krogstrup}}, \bibinfo {author} {\bibfnamefont {C.~M.}\ \bibnamefont
  {Marcus}}, \ and\ \bibinfo {author} {\bibfnamefont {Y.}~\bibnamefont
  {Oreg}},\ }\href {\doibase 10.1038/s41578-018-0003-1} {\bibfield  {journal}
  {\bibinfo  {journal} {Nat. Rev. Mater.}\ }\textbf {\bibinfo {volume} {3}},\
  \bibinfo {pages} {52} (\bibinfo {year} {2018})}\BibitemShut {NoStop}%
\bibitem [{\citenamefont {Drukier}\ \emph {et~al.}(2018)\citenamefont
  {Drukier}, \citenamefont {Zirnstein}, \citenamefont {Rosenow}, \citenamefont
  {Stern},\ and\ \citenamefont {Oreg}}]{PhysRevB.98.161401}%
  \BibitemOpen
  \bibfield  {author} {\bibinfo {author} {\bibfnamefont {C.}~\bibnamefont
  {Drukier}}, \bibinfo {author} {\bibfnamefont {H.-G.}\ \bibnamefont
  {Zirnstein}}, \bibinfo {author} {\bibfnamefont {B.}~\bibnamefont {Rosenow}},
  \bibinfo {author} {\bibfnamefont {A.}~\bibnamefont {Stern}}, \ and\ \bibinfo
  {author} {\bibfnamefont {Y.}~\bibnamefont {Oreg}},\ }\href {\doibase
  10.1103/PhysRevB.98.161401} {\bibfield  {journal} {\bibinfo  {journal} {Phys.
  Rev. B}\ }\textbf {\bibinfo {volume} {98}},\ \bibinfo {pages} {161401}
  (\bibinfo {year} {2018})}\BibitemShut {NoStop}%
\bibitem [{\citenamefont {Liu}\ \emph {et~al.}(2019)\citenamefont {Liu},
  \citenamefont {Cole},\ and\ \citenamefont {Sau}}]{PhysRevLett.122.117001}%
  \BibitemOpen
  \bibfield  {author} {\bibinfo {author} {\bibfnamefont {C.-X.}\ \bibnamefont
  {Liu}}, \bibinfo {author} {\bibfnamefont {W.~S.}\ \bibnamefont {Cole}}, \
  and\ \bibinfo {author} {\bibfnamefont {J.~D.}\ \bibnamefont {Sau}},\ }\href
  {\doibase 10.1103/PhysRevLett.122.117001} {\bibfield  {journal} {\bibinfo
  {journal} {Phys. Rev. Lett.}\ }\textbf {\bibinfo {volume} {122}},\ \bibinfo
  {pages} {117001} (\bibinfo {year} {2019})}\BibitemShut {NoStop}%
\bibitem [{\citenamefont {Hell}\ \emph {et~al.}(2018)\citenamefont {Hell},
  \citenamefont {Flensberg},\ and\ \citenamefont
  {Leijnse}}]{PhysRevB.97.161401}%
  \BibitemOpen
  \bibfield  {author} {\bibinfo {author} {\bibfnamefont {M.}~\bibnamefont
  {Hell}}, \bibinfo {author} {\bibfnamefont {K.}~\bibnamefont {Flensberg}}, \
  and\ \bibinfo {author} {\bibfnamefont {M.}~\bibnamefont {Leijnse}},\ }\href
  {\doibase 10.1103/PhysRevB.97.161401} {\bibfield  {journal} {\bibinfo
  {journal} {Phys. Rev. B}\ }\textbf {\bibinfo {volume} {97}},\ \bibinfo
  {pages} {161401} (\bibinfo {year} {2018})}\BibitemShut {NoStop}%
\bibitem [{\citenamefont {{Whiticar}}\ \emph {et~al.}(2019)\citenamefont
  {{Whiticar}}, \citenamefont {{Fornieri}}, \citenamefont {{O'Farrell}},
  \citenamefont {{Drachmann}}, \citenamefont {{Wang}}, \citenamefont
  {{Thomas}}, \citenamefont {{Gronin}}, \citenamefont {{Kallaher}},
  \citenamefont {{Gardner}}, \citenamefont {{Manfra}}, \citenamefont
  {{Marcus}},\ and\ \citenamefont {{Nichele}}}]{1902.07085}%
  \BibitemOpen
  \bibfield  {author} {\bibinfo {author} {\bibfnamefont {A.~M.}\ \bibnamefont
  {{Whiticar}}}, \bibinfo {author} {\bibfnamefont {A.}~\bibnamefont
  {{Fornieri}}}, \bibinfo {author} {\bibfnamefont {E.~C.~T.}\ \bibnamefont
  {{O'Farrell}}}, \bibinfo {author} {\bibfnamefont {A.~C.~C.}\ \bibnamefont
  {{Drachmann}}}, \bibinfo {author} {\bibfnamefont {T.}~\bibnamefont {{Wang}}},
  \bibinfo {author} {\bibfnamefont {C.}~\bibnamefont {{Thomas}}}, \bibinfo
  {author} {\bibfnamefont {S.}~\bibnamefont {{Gronin}}}, \bibinfo {author}
  {\bibfnamefont {R.}~\bibnamefont {{Kallaher}}}, \bibinfo {author}
  {\bibfnamefont {G.~C.}\ \bibnamefont {{Gardner}}}, \bibinfo {author}
  {\bibfnamefont {M.~J.}\ \bibnamefont {{Manfra}}}, \bibinfo {author}
  {\bibfnamefont {C.~M.}\ \bibnamefont {{Marcus}}}, \ and\ \bibinfo {author}
  {\bibfnamefont {F.}~\bibnamefont {{Nichele}}},\ }\href@noop {} {\bibfield
  {journal} {\bibinfo  {journal} {arXiv e-prints}\ ,\ \bibinfo {eid}
  {arXiv:1902.07085}} (\bibinfo {year} {2019})}\BibitemShut {NoStop}%
\bibitem [{\citenamefont {Kane}\ \emph {et~al.}(2017)\citenamefont {Kane},
  \citenamefont {Stern},\ and\ \citenamefont {Halperin}}]{PhysRevX.7.031009}%
  \BibitemOpen
  \bibfield  {author} {\bibinfo {author} {\bibfnamefont {C.~L.}\ \bibnamefont
  {Kane}}, \bibinfo {author} {\bibfnamefont {A.}~\bibnamefont {Stern}}, \ and\
  \bibinfo {author} {\bibfnamefont {B.~I.}\ \bibnamefont {Halperin}},\ }\href
  {\doibase 10.1103/PhysRevX.7.031009} {\bibfield  {journal} {\bibinfo
  {journal} {Phys. Rev. X}\ }\textbf {\bibinfo {volume} {7}},\ \bibinfo {pages}
  {031009} (\bibinfo {year} {2017})}\BibitemShut {NoStop}%
\bibitem [{\citenamefont {Manucharyan}\ \emph {et~al.}(2009)\citenamefont
  {Manucharyan}, \citenamefont {Koch}, \citenamefont {Glazman},\ and\
  \citenamefont {Devoret}}]{Science.326.113}%
  \BibitemOpen
  \bibfield  {author} {\bibinfo {author} {\bibfnamefont {V.~E.}\ \bibnamefont
  {Manucharyan}}, \bibinfo {author} {\bibfnamefont {J.}~\bibnamefont {Koch}},
  \bibinfo {author} {\bibfnamefont {L.~I.}\ \bibnamefont {Glazman}}, \ and\
  \bibinfo {author} {\bibfnamefont {M.~H.}\ \bibnamefont {Devoret}},\ }\href
  {\doibase 10.1126/science.1175552} {\bibfield  {journal} {\bibinfo  {journal}
  {Science}\ }\textbf {\bibinfo {volume} {326}},\ \bibinfo {pages} {113}
  (\bibinfo {year} {2009})}\BibitemShut {NoStop}%
\bibitem [{\citenamefont {Masluk}\ \emph {et~al.}(2012)\citenamefont {Masluk},
  \citenamefont {Pop}, \citenamefont {Kamal}, \citenamefont {Minev},\ and\
  \citenamefont {Devoret}}]{PhysRevLett.109.137002}%
  \BibitemOpen
  \bibfield  {author} {\bibinfo {author} {\bibfnamefont {N.~A.}\ \bibnamefont
  {Masluk}}, \bibinfo {author} {\bibfnamefont {I.~M.}\ \bibnamefont {Pop}},
  \bibinfo {author} {\bibfnamefont {A.}~\bibnamefont {Kamal}}, \bibinfo
  {author} {\bibfnamefont {Z.~K.}\ \bibnamefont {Minev}}, \ and\ \bibinfo
  {author} {\bibfnamefont {M.~H.}\ \bibnamefont {Devoret}},\ }\href {\doibase
  10.1103/PhysRevLett.109.137002} {\bibfield  {journal} {\bibinfo  {journal}
  {Phys. Rev. Lett.}\ }\textbf {\bibinfo {volume} {109}},\ \bibinfo {pages}
  {137002} (\bibinfo {year} {2012})}\BibitemShut {NoStop}%
\bibitem [{\citenamefont {Bell}\ \emph {et~al.}(2012)\citenamefont {Bell},
  \citenamefont {Sadovskyy}, \citenamefont {Ioffe}, \citenamefont {Kitaev},\
  and\ \citenamefont {Gershenson}}]{PhysRevLett.109.137003}%
  \BibitemOpen
  \bibfield  {author} {\bibinfo {author} {\bibfnamefont {M.~T.}\ \bibnamefont
  {Bell}}, \bibinfo {author} {\bibfnamefont {I.~A.}\ \bibnamefont {Sadovskyy}},
  \bibinfo {author} {\bibfnamefont {L.~B.}\ \bibnamefont {Ioffe}}, \bibinfo
  {author} {\bibfnamefont {A.~Y.}\ \bibnamefont {Kitaev}}, \ and\ \bibinfo
  {author} {\bibfnamefont {M.~E.}\ \bibnamefont {Gershenson}},\ }\href
  {\doibase 10.1103/PhysRevLett.109.137003} {\bibfield  {journal} {\bibinfo
  {journal} {Phys. Rev. Lett.}\ }\textbf {\bibinfo {volume} {109}},\ \bibinfo
  {pages} {137003} (\bibinfo {year} {2012})}\BibitemShut {NoStop}%
\bibitem [{\citenamefont {Kuzmin}\ \emph {et~al.}(2019)\citenamefont {Kuzmin},
  \citenamefont {Mencia}, \citenamefont {Grabon}, \citenamefont {Mehta},
  \citenamefont {Lin},\ and\ \citenamefont {Manucharyan}}]{NatPhys.15.930}%
  \BibitemOpen
  \bibfield  {author} {\bibinfo {author} {\bibfnamefont {R.}~\bibnamefont
  {Kuzmin}}, \bibinfo {author} {\bibfnamefont {R.}~\bibnamefont {Mencia}},
  \bibinfo {author} {\bibfnamefont {N.}~\bibnamefont {Grabon}}, \bibinfo
  {author} {\bibfnamefont {N.}~\bibnamefont {Mehta}}, \bibinfo {author}
  {\bibfnamefont {Y.-H.}\ \bibnamefont {Lin}}, \ and\ \bibinfo {author}
  {\bibfnamefont {V.~E.}\ \bibnamefont {Manucharyan}},\ }\href {\doibase
  10.1038/s41567-019-0553-1} {\bibfield  {journal} {\bibinfo  {journal} {Nat.
  Phys.}\ }\textbf {\bibinfo {volume} {15}},\ \bibinfo {pages} {930} (\bibinfo
  {year} {2019})}\BibitemShut {NoStop}%
\bibitem [{\citenamefont {Sau}\ \emph {et~al.}(2011)\citenamefont {Sau},
  \citenamefont {Halperin}, \citenamefont {Flensberg},\ and\ \citenamefont
  {Das~Sarma}}]{PhysRevB.84.144509}%
  \BibitemOpen
  \bibfield  {author} {\bibinfo {author} {\bibfnamefont {J.~D.}\ \bibnamefont
  {Sau}}, \bibinfo {author} {\bibfnamefont {B.~I.}\ \bibnamefont {Halperin}},
  \bibinfo {author} {\bibfnamefont {K.}~\bibnamefont {Flensberg}}, \ and\
  \bibinfo {author} {\bibfnamefont {S.}~\bibnamefont {Das~Sarma}},\ }\href
  {\doibase 10.1103/PhysRevB.84.144509} {\bibfield  {journal} {\bibinfo
  {journal} {Phys. Rev. B}\ }\textbf {\bibinfo {volume} {84}},\ \bibinfo
  {pages} {144509} (\bibinfo {year} {2011})}\BibitemShut {NoStop}%
\bibitem [{\citenamefont {Knapp}\ \emph {et~al.}(2020)\citenamefont {Knapp},
  \citenamefont {V\"ayrynen},\ and\ \citenamefont
  {Lutchyn}}]{PhysRevB.101.125108}%
  \BibitemOpen
  \bibfield  {author} {\bibinfo {author} {\bibfnamefont {C.}~\bibnamefont
  {Knapp}}, \bibinfo {author} {\bibfnamefont {J.~I.}\ \bibnamefont
  {V\"ayrynen}}, \ and\ \bibinfo {author} {\bibfnamefont {R.~M.}\ \bibnamefont
  {Lutchyn}},\ }\href {\doibase 10.1103/PhysRevB.101.125108} {\bibfield
  {journal} {\bibinfo  {journal} {Phys. Rev. B}\ }\textbf {\bibinfo {volume}
  {101}},\ \bibinfo {pages} {125108} (\bibinfo {year} {2020})}\BibitemShut
  {NoStop}%
\bibitem [{Note1()}]{Note1}%
  \BibitemOpen
  \bibinfo {note} {See Supplemental Material at [URL will be inserted by
  publisher] for the derivation of Eq.\ (\ref {eq:DeltaG}).}\BibitemShut
  {Stop}%
\bibitem [{\citenamefont {Roy}\ \emph {et~al.}(2020)\citenamefont {Roy},
  \citenamefont {Hauschild},\ and\ \citenamefont
  {Pollmann}}]{PhysRevB.101.075419}%
  \BibitemOpen
  \bibfield  {author} {\bibinfo {author} {\bibfnamefont {A.}~\bibnamefont
  {Roy}}, \bibinfo {author} {\bibfnamefont {J.}~\bibnamefont {Hauschild}}, \
  and\ \bibinfo {author} {\bibfnamefont {F.}~\bibnamefont {Pollmann}},\ }\href
  {\doibase 10.1103/PhysRevB.101.075419} {\bibfield  {journal} {\bibinfo
  {journal} {Phys. Rev. B}\ }\textbf {\bibinfo {volume} {101}},\ \bibinfo
  {pages} {075419} (\bibinfo {year} {2020})}\BibitemShut {NoStop}%
\bibitem [{\citenamefont {Houzet}\ and\ \citenamefont
  {Glazman}(2019)}]{PhysRevLett.122.237701}%
  \BibitemOpen
  \bibfield  {author} {\bibinfo {author} {\bibfnamefont {M.}~\bibnamefont
  {Houzet}}\ and\ \bibinfo {author} {\bibfnamefont {L.~I.}\ \bibnamefont
  {Glazman}},\ }\href {\doibase 10.1103/PhysRevLett.122.237701} {\bibfield
  {journal} {\bibinfo  {journal} {Phys. Rev. Lett.}\ }\textbf {\bibinfo
  {volume} {122}},\ \bibinfo {pages} {237701} (\bibinfo {year}
  {2019})}\BibitemShut {NoStop}%
\bibitem [{\citenamefont {Devoret}\ \emph {et~al.}(2014)\citenamefont
  {Devoret}, \citenamefont {Huard}, \citenamefont {Schoelkopf},\ and\
  \citenamefont {Cugliandolo}}]{devoret2014quantum}%
  \BibitemOpen
  \bibfield  {author} {\bibinfo {author} {\bibfnamefont {M.}~\bibnamefont
  {Devoret}}, \bibinfo {author} {\bibfnamefont {B.}~\bibnamefont {Huard}},
  \bibinfo {author} {\bibfnamefont {R.}~\bibnamefont {Schoelkopf}}, \ and\
  \bibinfo {author} {\bibfnamefont {L.}~\bibnamefont {Cugliandolo}},\ }\href
  {https://books.google.de/books?id=FiznAwAAQBAJ} {\emph {\bibinfo {title}
  {Quantum Machines: Measurement and Control of Engineered Quantum Systems}}},\
  Ecole De Physique Des Houches\ (\bibinfo  {publisher} {Oxford University
  Press},\ \bibinfo {year} {2014})\BibitemShut {NoStop}%
\bibitem [{\citenamefont {Nazarov}(1989{\natexlab{a}})}]{ZhEkspTeorFiz.95.975}%
  \BibitemOpen
  \bibfield  {author} {\bibinfo {author} {\bibfnamefont {Y.~V.}\ \bibnamefont
  {Nazarov}},\ }\href@noop {} {\bibfield  {journal} {\bibinfo  {journal} {Zh.
  Eksp. Teor. Fiz.}\ }\textbf {\bibinfo {volume} {95}},\ \bibinfo {pages} {975}
  (\bibinfo {year} {1989}{\natexlab{a}})}\BibitemShut {NoStop}%
\bibitem [{\citenamefont {Nazarov}(1989{\natexlab{b}})}]{SovPhysJETP.68.561}%
  \BibitemOpen
  \bibfield  {author} {\bibinfo {author} {\bibfnamefont {Y.~V.}\ \bibnamefont
  {Nazarov}},\ }\href@noop {} {\bibfield  {journal} {\bibinfo  {journal} {Sov.
  Phys. JETP}\ }\textbf {\bibinfo {volume} {68}},\ \bibinfo {pages} {561}
  (\bibinfo {year} {1989}{\natexlab{b}})}\BibitemShut {NoStop}%
\bibitem [{\citenamefont
  {Nazarov}(1989{\natexlab{c}})}]{PismaZhEkspTeorFiz.49.105}%
  \BibitemOpen
  \bibfield  {author} {\bibinfo {author} {\bibfnamefont {Y.~V.}\ \bibnamefont
  {Nazarov}},\ }\href@noop {} {\bibfield  {journal} {\bibinfo  {journal}
  {Pis'ma Zh. Eksp. Teor. Fiz.}\ }\textbf {\bibinfo {volume} {49}},\ \bibinfo
  {pages} {105} (\bibinfo {year} {1989}{\natexlab{c}})}\BibitemShut {NoStop}%
\bibitem [{\citenamefont {Nazarov}(1989{\natexlab{d}})}]{JETPLett.49.126}%
  \BibitemOpen
  \bibfield  {author} {\bibinfo {author} {\bibfnamefont {Y.~V.}\ \bibnamefont
  {Nazarov}},\ }\href@noop {} {\bibfield  {journal} {\bibinfo  {journal} {JETP
  Lett.}\ }\textbf {\bibinfo {volume} {49}},\ \bibinfo {pages} {126} (\bibinfo
  {year} {1989}{\natexlab{d}})}\BibitemShut {NoStop}%
\bibitem [{\citenamefont {Devoret}\ \emph {et~al.}(1990)\citenamefont
  {Devoret}, \citenamefont {Esteve}, \citenamefont {Grabert}, \citenamefont
  {Ingold}, \citenamefont {Pothier},\ and\ \citenamefont
  {Urbina}}]{PhysRevLett.64.1824}%
  \BibitemOpen
  \bibfield  {author} {\bibinfo {author} {\bibfnamefont {M.~H.}\ \bibnamefont
  {Devoret}}, \bibinfo {author} {\bibfnamefont {D.}~\bibnamefont {Esteve}},
  \bibinfo {author} {\bibfnamefont {H.}~\bibnamefont {Grabert}}, \bibinfo
  {author} {\bibfnamefont {G.-L.}\ \bibnamefont {Ingold}}, \bibinfo {author}
  {\bibfnamefont {H.}~\bibnamefont {Pothier}}, \ and\ \bibinfo {author}
  {\bibfnamefont {C.}~\bibnamefont {Urbina}},\ }\href {\doibase
  10.1103/PhysRevLett.64.1824} {\bibfield  {journal} {\bibinfo  {journal}
  {Phys. Rev. Lett.}\ }\textbf {\bibinfo {volume} {64}},\ \bibinfo {pages}
  {1824} (\bibinfo {year} {1990})}\BibitemShut {NoStop}%
\bibitem [{\citenamefont {Girvin}\ \emph {et~al.}(1990)\citenamefont {Girvin},
  \citenamefont {Glazman}, \citenamefont {Jonson}, \citenamefont {Penn},\ and\
  \citenamefont {Stiles}}]{PhysRevLett.64.3183}%
  \BibitemOpen
  \bibfield  {author} {\bibinfo {author} {\bibfnamefont {S.~M.}\ \bibnamefont
  {Girvin}}, \bibinfo {author} {\bibfnamefont {L.~I.}\ \bibnamefont {Glazman}},
  \bibinfo {author} {\bibfnamefont {M.}~\bibnamefont {Jonson}}, \bibinfo
  {author} {\bibfnamefont {D.~R.}\ \bibnamefont {Penn}}, \ and\ \bibinfo
  {author} {\bibfnamefont {M.~D.}\ \bibnamefont {Stiles}},\ }\href {\doibase
  10.1103/PhysRevLett.64.3183} {\bibfield  {journal} {\bibinfo  {journal}
  {Phys. Rev. Lett.}\ }\textbf {\bibinfo {volume} {64}},\ \bibinfo {pages}
  {3183} (\bibinfo {year} {1990})}\BibitemShut {NoStop}%
\end{thebibliography}%

\end{document}